# Measuring Cost of Quality (CoQ) on SDLC Projects is Indispensible for Effective Software Quality Assurance

[*1] Parvez Mahmood Khan, [2] M.M. Sufyan Beg
Department of Computer Engineering, J.M.I. New Delhi, India
[1] pmkhan@hotmail.com , [2] mmsbeg@cs.berkeley.edu

*Abstract* It is well known fact that was phrased by famous quality scholar P.B. Crosby that "*it is always cheaper to do the job right the first time*". However, this statement must be reconsidered with respect to software development projects, because the concept of quality and associated costs measurements in software engineering discipline is not as matured as in manufacturing and other fields of the industry. Post delivery defects (i.e. software bugs) are very common and integral part of software industry. While the process of measuring and classifying quality cost components is visible, obvious and institutionalized in manufacturing industry, it is still evolving in software industry. In addition to this, the recommendations of British standard BS-6143-2:1990 for classifying quality-related costs into prevention costs, appraisal costs, and failure costs have been successfully adopted by many industries, by identifying the activities carried out within each of these categories, and measuring the costs connected with them, software industry has a long-way to go to have the same level of adoption and institutionalization of cost of quality measurements and visibility. *Cost of Quality* for software isn't the price of creating a quality software product or IT-service. It's actually the cost of NOT creating a quality software product or IT-service. The chronic affliction of majority of software development projects that are frequently found bleeding with *cost overruns*, *schedule slippage*, *scope creep* and ***poor quality of deliverables*** in the global IT industry, was the trigger for this research work. The idea was to examine a good number of SDLC-projects (proper mix of successful projects as well as failed projects) from multiple organizations end-to-end (i.e. from project-inception to project-closure) and empirically assess the quality management approach – focusing on how the quality was planned on these project(s), what specific *software quality assurance* and *software quality control* measured were employed on the projects under study and it's possible impact on overall project success and achievement of business objectives. Lessons learnt from this study offer valuable prescriptive guidance for small and medium software businesses, who can benefit from this study by applying the same for their quality improvement initiatives using CoQ-metric, to enhance the capability and maturity of their SDLC-project performance.

* Corresponding address:
Parvez Mahmood Khan,
Department of Computer Engineering, J.M.I. New Delhi, India
[1] pmkhan@hotmail.com

## 1. Introduction

Software development projects are very often characterized by severe *cost overruns*, *schedule slippages* and **poor quality** of deliverables. In order to meet the quality expectations of customers, lot of time and effort is spent by software seller organizations on fire fighting after the software product is released (attending to failures/ fixing critical bugs etc.) which has a cost implication. Unfortunately, these costs are not clearly understood. These costs often disappear as the costs of testing, the general developments costs, or the operating expenses which is misleading. In the context of recent global recession, when each and every organization is looking at ways and means of controlling and reducing the costs, this study has been undertaken to help understand the actual components of software quality costs – with view of providing better visibility to the software development organizations, thereby helping them improve their cost effectiveness and consequent financial health of the organization. For our current study, we have looked at analogy of quality costs in manufacturing domain and gathered that for manufacturing, the British standard BS-6143-2:1990 classifies quality-related costs into prevention costs, appraisal costs, and failure costs. It furthermore recommends to identify the activities carried out within each of these categories, and to measure the costs connected with the activities. The British standard BS-6143-2:1990 thus presents a framework for recording and structuring costs once they have occurred. We have therefore, used the ideas from  BS-6142-2:1990 framework for our current study, with a view of apply it's concepts on software development projects.

In this research work, an effort has been made to study and evaluate the impact of Quality management activities on software development project – using cost of quality as a metric, across a good number of real-life SDLC-projects, over multiple organizations (SMBs as well as large organizations).

## 2.   Basic Overview of Quality Management

Literature survey on the definitions of quality has shown a great deal of variation. Some of the commonly referred definitions of quality for software products/IT-Services are as follows:

*Webster's Definition of Quality*:
The totality of features and characteristics of a product or service that bear on its ability to satisfy stated or implied needs[1].

*Crosby's view of Quality* –*Management's Perspective:*
Software quality is **conformance** of the software solution (be it software product or IT- service) **to the requirements** [2, 3].

*Juran's view of Quality* –*Manufacturer's Perspective:*
Software quality is **fitness** of the software product (or IT Service) **for  intended use** [4].

*Deming's view of Quality* – *Consumer's Perspective:*
Software quality is **ability** of the software product (or IT-Service) **to meet** or exceed **the end-user expectations**[5].

It is apparent from a first look at these definitions that quality is a subjective term with multitude of definitions – covering different perspectives of various stakeholders. However, a closer look and analysis of these definitions from the perspective of intent reveals that while all the three major quality leaders  (Croby, Juran & Deming) have their own ideas on how the quality should be measured and managed in the organizations, but they are all essentially pointing in the same direction – "quality is non-negotiable". It is therefore important, for the organizations - to manage quality using a defined and structured approach to quality management.

In general, *Project Quality Management* on any project is meant to address two aspects: the management of the *project itself* as well as management of the *product of the project*. While the project aspects are common and independent of the type of project, but the product of the project aspects vary greatly depending upon the domain concerned. For example, quality management of software product produced by an SDLC-project would be very different from quality management of a housing apartment produced by a construction project, even though the project aspects of quality management of both SDLC-project and construction-project will be common. The objective of quality management on the software development projects is to ensure that quality objectives of the software project deliverables are achieved and cost of quality is reduced (by minimizing post delivery defects on deliverables of the SDLC-project). Improved quality that exceeds customer expectations will generate more revenues that exceed the cost of quality.

There are many *proprietary approaches* to quality management such as those recommended by Deming, Juran, Crosby and many other quality pundits and several *non proprietary approaches* like Total Quality Management(TQM), Failure Mode and Effect Analysis(FEMA), Design reviews, Voice of customer(VOC), Cost of Quality(COQ), etc.

In our study of various software development organizations, we have found that organizations adopt and maintain their QMS(Quality Management System) – based on a combination of propriety and non-propriety approaches, as per the capability and maturity of the organization and its quality focus. There is no single best way to deploy a Quality Management System(QMS) in the organization, although having an organizational QMS has been found to help quality management on software development projects of the organization concerned. The objective of quality management on the software development projects is to ensure that *quality objectives* of the project deliverables are achieved and *cost of quality* is reduced (by minimizing post delivery defects on deliverables of the SDLC-project). At the project level, the <u>Quality Management Knowledge Area</u>[6] spreads across the following 3-processes, which can be easily tailored in-line with organizational QMS, for adoption on the software development projects:

- · *Plan Quality* – The process of identifying quality requirements and/or standards for the project and product, and documenting how the project will demonstrate compliance.

- · *Perform Quality Assurance* – The process of auditing the quality requirements and the results from quality control measures to ensure appropriate quality standards are used.

- · *Perform Quality Control* – The process of monitoring and recording results of executing the quality activities to assess performance and recommend necessary changes.

One of the fundamental tenets of modern quality management is that quality should be planned, designed and built in – not inspected in. This is because the cost of preventing mistakes is much less than the cost of correcting them when they are found by inspection. Consequently, quality conscious organizations are striving to have a shift in the organizational culture from Quality Control to Quality Assurance. Quality Assurance(QA) is more effective than Quality Control(QC) – because in QA, the emphasis moves to the development process. This enables attempt to fix problems before and during the development process itself. Lessons learnt from past projects help to improve the process and therefore reduce the number of defects in a lasting manner. While institutionalizing the software quality assurance processes, there is a need to implement a system of methods and procedures that are typically used to assure that the software products delivered as a result of SDLC-projects are essentially meeting the requirements. These methods and procedures, typically include planning, measuring and monitoring of all the works[6],[7] performed by project team (including software engineers, software testers etc.).

## 3. Research Background & Context

This study was undertaken with real world projects data from seven different business organizations, to evaluate the methods & practices of quality assurance actually used and practiced on software development projects selected for this study (from businesses organizations of IT-Industry in India). The focus was on working towards a better understanding of quality management processes that are actually practiced in real world projects encompassing software development. The overall objective of the study is to propose better methods and prescriptive guidance for planning and execution of software development projects that can and will help successful delivery of "quality products" in a cost effective manner, from business critical software project management.

## 4. Observations on Related Work

A good overview of quality fundamentals and basic concepts is available in literature [1]-[5]. Although many sectors (like Automobile, Aviation, Construction etc.) have really implemented and institutionalized robust quality processes, but adoption of quality processes within software engineering domain is still a challenge. Software development Projects are very intricate and risky endaviours requiring careful integration of various disciplines, technical activities, project management etc. Managing quality of deliverables from a software project has been an area of concern for quite some-time, but it is, gaining much more interest in recent times due to the various factors economic, social and fierce competition among software vendors [17]. An unprecedented emphasis is being associated these days to the production of high quality software products [18]. In [8], Nasib S. Gill described how insecurely tested software system lowers down the system reliability that afterward negatively affects 'Software Quality'. Also 'Software Reliability Measurement' has been discussed and along-with ISO approach applicable to software quality assurance (SQA). In order to increase the efficiency of testing and to improve software quality, software houses must make transitions to higher software culture. It is suggested that testing needs to concentrate on maximizing customer satisfaction rather than just detecting and correcting errors involved in delivered software. Author also discussed the factors affecting software quality management and suggested possible improvements. The results of this paper may be useful and supportive to the researchers in quantifying the specific measuring tools for these software qualities attributes.

In [9] L H Rosenberg.et.al articulated the fact that software quality assurance is faced with many challenges starting with the method of defining quality for software. On one hand there is a need to have complete understanding of what high quality software is, but on the other hand the final description is generally influenced by the environment of the software usage. Also, there are many aspects of SQA from those within the phases of the software development life cycle to those that span several phases. SQA is a very difficult area that can greatly influence the final outcome of a project. It is also one that requires a rather diverse set of skills. New information areas such as software safety and reliability are now being added to the core set of required skills. SQA must be independent from development organizations to be successful.

In [10] M. Agrawal.Et.al elaborated on their work pertaining to understanding the implication of process maturity on the goals of developing high-quality software on-time and within budget - by specifically focusing only CMM Level 5 projects from industry, spanning multiple organizations and projects. The goal was to study the impacts of highly mature processes on effort, quality, and cycle time. They used a linear regression model based on data collected from 37 CMM level 5 projects of four organizations. They found that high levels of process maturity, as indicated by CMM level 5 rating, reduce the effects of most factors that were previously believed to impact software development effort, quality, and cycle time. The only factor that they found to be significant in determining effort, cycle time, and quality was software size. On the average, the developed models predicted effort and cycle time around 12 percent and defects to about 49 percent of the actual, across organizations. Overall, the results of their study indicates that some of the biggest rewards from high levels of process maturity come from the reduction in variance of software development outcomes that were caused by factors other than software size.

In [12] A. Khanjani.et.al articulated quality assurance issues and challenges under the context of open source software development (OSSD) process. Open Source Software (OSS) is software products available to the public, with its source code. Anyone can study, change, and improve the design of open source products.

Accordingly, Open Source Software Development (OSSD) is the process by which open source software is developed within the confines of software engineering life-cycle methods. However when open source used for commercial purpose, then an open source license is required. Although Open source software is generally developed in a public and collaborative manner, but there are defined quality assurance principles under open source software development with a view to improve software product quality against traditional methods and techniques. Despite wide spread acceptance and adoption of the open source developments in recent years, there are a number of product quality issues and challenges facing the open source development model. Many industries and business sectors are following or using OSSD, since they realize the benefits, but they do have some reservations concerning quality assurance in the form of program code quality, maintenance of the code and its quality, over the life-cycle of the product and third party usage. This paper has reviewed the literature of the process of the latest quality assurance, under open source software development methods and techniques. The result from this review is to demonstrate the process of quality assurance of open source software and that how it can affect the overall quality assurance principles.

In [13] M.K. Omar.et.al have examined an existing cost of quality model and modified the same in the context of extension of previous work. The modified model is then used to develop a simulation model for two processing stations in a manufacturing system. Using real life data from the industry and through simulation works, the model has been tested with a set of managerial quality control decision scenarios which are commonly practiced in real life industry. The results shown are interpreted to be that some of the managerial quality control decision scenarios have significant impact on cost of quality.

In [14] C. Woody.et.al have emphasized the need for effective software quality assurance in the wake of our society's growing dependence on software products. Need for software quality assurance, understanding of what to do and how to address actions required for quality assurance is articulated in the paper. Efforts currently underway for disseminating the principles for software quality assurance and how to educate those involved in the process is also elaborated in this paper.

In [15] A. Janus.et.al have proposed a new approach for agile quality assurance by combining the concept of software metrics of quality measurements with agile practice of continuous integration. The proposed method adds continuous measurement and continuous improvement as subsequent activities to CI and establishes Metric-based Quality-Gates for an Agile Quality Assurance. This method has been deployed in an Agile Maintenance and Evolution project for the Automotive Industry at T-Systems International - a large German ICT company and found working satisfactorily with a project environment that used a (legacy) Java-based Web Application including the use of Open Source Tools from the Java Eco-System.

In [16], Wei Li has investigated the problem for component-based software systems from three points of view. First, the whole spectrum of QoS (Quality of Service) characteristics is defined. Second, the logical and physical requirements for QoS characteristics are analyzed and solutions to achieve them are proposed. Third, prior work is classified by QoS characteristics and then realized by abstract reconfiguration strategies. Subsequently, a quantitative evaluation of the QoS assurance abilities is carried out and the results are discussed. The paper concluded that classified QoS characteristics are achievable within the acceptable limits under practical constraints.

In [11] Manar Abu Talib.Et.al provided an overview of quantitative analysis techniques for software quality and their applicability during the software development life cycle (SDLC). The details included the Seven Basic Tools of Quality, Statistical Process Control, and Six Sigma, with special focus on how these techniques can be used for managing and controlling the quality of software during specification, design, implementation, testing, and maintenance. The paper also tried a verification of whether or not these techniques, which are generally accepted for most projects, most of the time, and have value that is recognized by the peer community, have indeed been included in the Guide to Software Engineering Body of Knowledge.

As noted above, despite these research advancements, empirical evidence suggests that quality is still a problem because developers lack an understanding of the source of problems, have an inability to learn from mistakes, lack effective tools, and do not have a complete verification process [7]. Also, there is a very clear need and direction to focus more on "quality assurance" and complement it with "quality control" in order to manage quality cost-effectively, particularly in the context of recent cycles of global recession and economic slowdowns. Also, very little research work has been devoted specifically to measuring cost of quality for software products, despite growing trend of costs and decline in quality of software products.

## 5. Research Methodology

Grounded theory research methods were adopted for this research for two reasons. First, the research was aimed at the extension of existing theory. Grounded theory is generally deemed appropriate for such efforts as it allows theory to emerge from interview transcripts and organization artifacts [7]. Second, the methodology is reputed to help separate researcher biases from interpretation of the data. A total of 28-senior leadership team members (covering Chief Quality Officers/CEOs, CFOs/Program Managers, CIOs/PMO Heads and Project Managers) from 7-different organizations were interviewed over a period of 12-months. The project managers chosen for the interview were generally very experienced in their position with average experience of at least 3-project deliveries end-to-end and had managed project budgets of multimillion- dollors. Four of the executives were retirees, having left the organization within the preceding couple of years. The interviews were facilitated via open-ended questions intended to encourage individuals to describe their needs and expectations from Quality Management System. Interviews also had discussions related to CoQ components and possibility of a correlation of CoQ-components with revenues and gross margins generated by business critical projects. The resulting pages of interview transcripts were analysed to identify recurrent themes and concepts associated with good and bad times in the business. Those themes and concepts were then tested against archived proejct's data including Project Data and Metrics Reports (PDMRs), transactional data from Project Management Information System (PMIS) and monthly progress reports etc. with an eye to find evidence reinforcing or challenging the themes and concepts. Preliminary conclusions were then tested via follow-up interviews with a subset of original 28-individuals. During the course of this study, the **actual processes employed on the projects were compared against an standard process of *plan quality*, *perform quality assurance* and *perform quality control* processes advocated by PMBOK[6].** DFDs for these standard processes used in our study as baselines for reference and analysis[6] are included in Figure-1, Figure-2 and Figure-3 respectively.

### 5.1 Plan Quality:

Standard Process for Planning Quality is depicted below in Figure-1. Planning for quality includes the process of identifying quality requirements and/or standards for the project itself as well as for the product of the project. It also includes documenting how the project will demonstrate compliance.

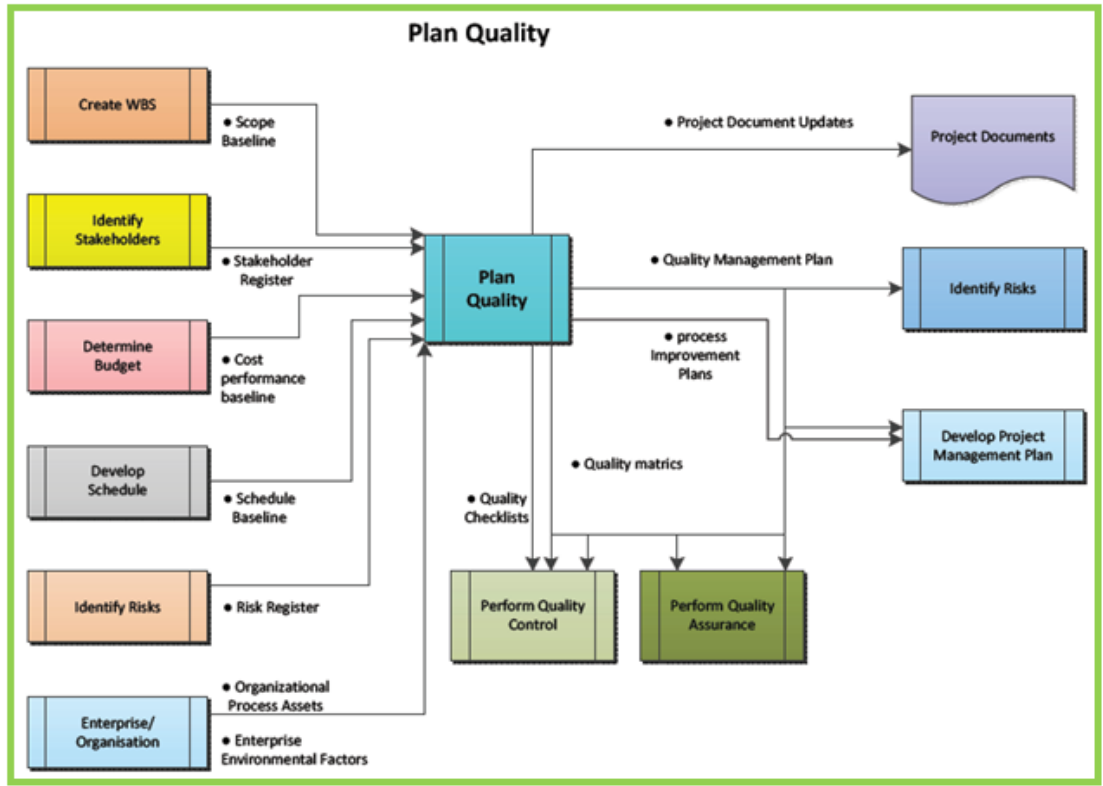

: *DFD for Plan Quality Process on SDLC- Projects*

## 5.2 Perform Quality Assurance:

Perform Quality Assurance process (depicted here in Fig-2) is the process of auditing quality requirements and the results from quality control measurements to ensure appropriate quality standards and used and operational definitions are consistently used across the project(s). Proper implementation of "Perform Quality Assurance Process" on a SDLC-project involves many things – like selecting the right people, engaging them in the correct activities, and providing appropriate support tools. It is driven by a common understanding that quality must be designed and built into software products/solutions, is everyone's responsibility, and is derived from a project's overall design. A successful Software Quality Assurance process advocates reviews in every phase of the SDLC and iterative build cycles to address quality issues. By focusing on defect prevention (as against defect detection and correction) from the start of a project, significant reductions in the total effort can be achieved through the avoidance of the rework/thrashing stage that is typically found on so many software development projects.

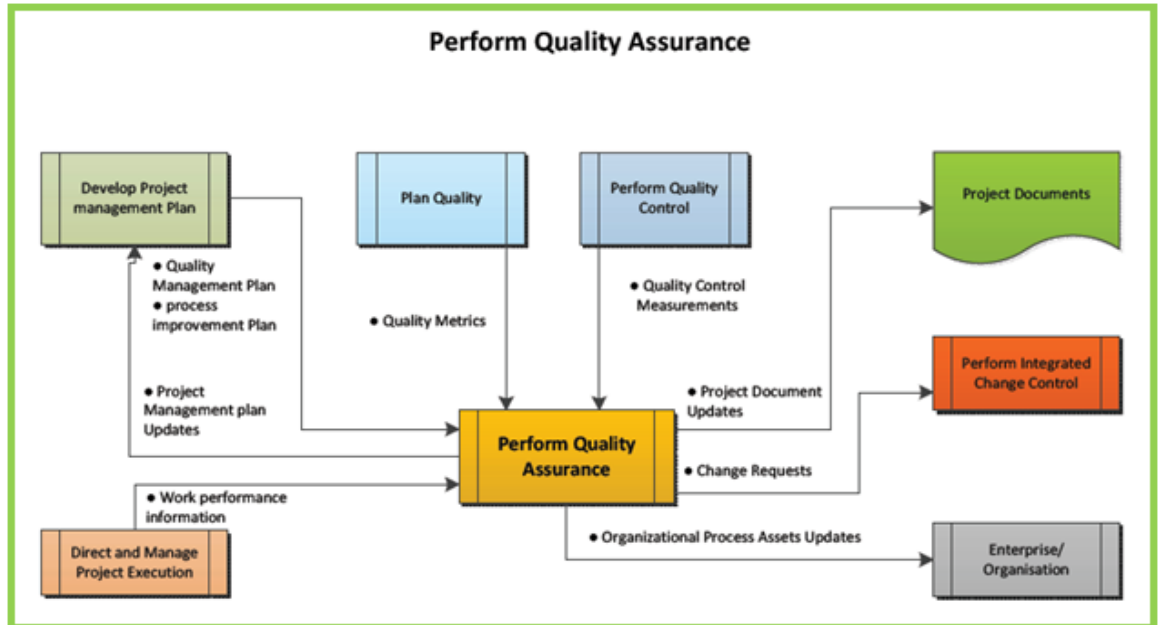

**Figure-2**: _DFD for Perform Quality Assurance Process for Software Projects_

## 5.3 Perform Quality Control:

Perform quality control process (depicted in Figure-3) is the process of monitoring and recording results of executing the quality activities to assess performance and recommend necessary changes.

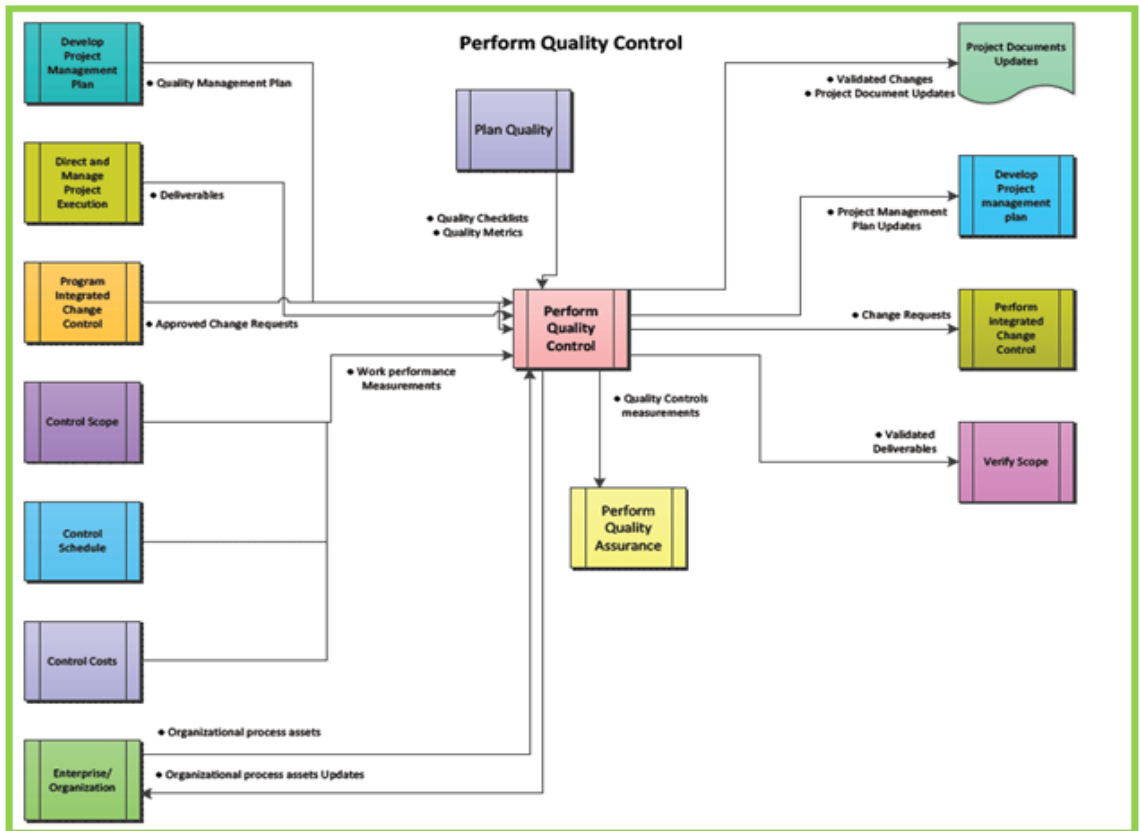

**Figure-3**: *DFD for Perform Quality Control Process for Software Projects*

This iterative process (interviews, concepts identification, concepts aggregation, theme analysis, testing against organizational artifacts, review with interviewees, adaption, repeat) led to the identification of key concepts and vital lessons learnt, for CoQ as a metric for effective quality assurance on business critical software projects. Our most important learning experiences from this study are articulated in Section 7 and 8.

## 6. Role of Cost of Quality (CoQ) Metrics for Software Quality Assurance

Cost of Quality(CoQ) is a financial measure of the quality performance of the project organization. It is actually a measure of "*cost of lack of quality*" and can be viewed as cost of bad quality[6]. Understanding the CoQ helps the performing organizations to develop quality conformance as a business strategy and unique selling point to improve their products, services and ultimately the brand image. The Cost of Quality, is all the costs that occur beyond the cost of producing the product "right the first time". It includes the additional costs associated with assuring that the product delivered meets the quality goals established for the software product.

### 6.1 Components of Software Cost of Quality on SDLC-Projects:

The three categories of costs associated with producing quality software products, from a typical SDLC-project are as summarized below in Table-1:

**Table-1**: Summary of Software Cost of Quality Component

| | |
|---|---|
| 1 | **External Failure Cost**<br>Costs associated with defects found after the customer receives the product or service.<br>**For example:** *processing customer complaints, customer returns, warranty claims, product recalls.* |
| 2 | **Internal Failure Cost**<br>Costs associated with defects found before the customer receives the product or service.<br>**For example**: *scrap, rework, re-inspection, re-testing, material review, material downgrades.* |
| 3 | **Inspection (Appraisal) Cost**<br>Costs incurred to determine the degree of conformance to quality requirements (measuring, evaluating or auditing).<br>**For example:** *inspection, testing, process or service audits, calibration of measuring and test equipment.* |
| 4 | **Prevention Cost**<br>Costs incurred to prevent (keep failure and appraisal cost to a minimum) poor quality.<br>**For example:** *New Product Review, Quality Panning, Supplier Surveys, Process Reviews, Quality Improvement Teams, Education and Training.* |

### 6.1.1 Prevention Costs

Money required to prevent errors and to do the job right the first time. These normally require up-front costs for benefits that will be derived months or even years later. This category includes money spent on establishing methods and procedures, training workers, acquiring tools, and planning for quality. Prevention money is all spent before the product is actually built.

### 6.1.2 Appraisal Costs

Money spent to review completed products against requirements. Appraisal includes the cost of inspections, testing, and reviews. This money is spent after the product is built but before it is shipped to the user or moved into production.

### 6.1.3 Failure Costs

All costs associated with defective products that have been delivered to the user or moved into production. Some failure costs involve repairing products to make them meet requirements. Others are costs generated by failures such as the cost of operating faulty products, damage incurred by using them, and the costs associated with operating a Help Desk. Failure costs are also sometimes referred to as Non-conformance costs, and consists of two components: Internal Failures and External Failures. Accordingly we have two types of failure costs:

#### 6.1.3.1 Internal failure costs:

The costs of internal failure include all expenses that arise when test cases fail the first time they're run, as they often do. In general, a programmer incurs a cost of internal failure while debugging problems found during his/her own unit and component testing. Once the code it put into formal testing (generally by an independent test team), the costs of internal failure increase. During the process of testing, the test team carries out testing and reports the failure, the programmer(s) find and fix the fault, the release engineer produces a new release, the system administration team installs that release in the test environment, and the tester retests the new release to confirm the fix and to check for regression, etc.

#### 6.1.3.2 External failure costs:

The costs of external failure are those incurred when, rather than a tester finding a bug, the customer does. These costs will be even higher than those associated with either kind of internal failure, programmer-found or tester-found. In these cases, not only does the same process described for tester-found bugs occur, but you also incur the technical support overhead and the more expensive process of releasing a fix to the field rather than to the test lab. In addition, consider the intangible costs: Angry customers, damage to the company image, lost business, and maybe even be legal claims for liquidated damages etc.

IT organizations across the globe are facing common challenges, such as ever increasing customer expectations, reducing budgets and reduced time to market[8]. Undertaking CoQ measurements and analysis programs for SDLC-projects as a part of this research has shown interesting trends that is in-line with industry benchmarks. Exhibits-1& 2 are meant to provide better clarity on the CoQ-metric and it's constituents – the understanding of which is essential for establishing a quality management strategy for the organization.

## 6.2 Exhibit-1: *An example of CoQ Measurement*

For example, consider the time-sheet as shown below, to understand how we compute CoQ on SDLC- projects:

| Sr. No. | Project-Activity | Hours Spent | Is part of CoQ? | Component of CoQ |
|---|---|---|---|---|
| | Let us consider the time-sheet of a typical project, with efforts (in hours). We need to compute **CoQ** (Cost of Quality) in hours and classify the components of CoQ into *Prevention*, *Appraisal* & *Failures* categories | | | |
| 1 | Training | 10-hours | Yes | Prevention Cost |
| 2 | Requirements Gathering | 25-hours | No | Not Applicable |
| 3 | Requirements Review | 5-hours | Yes | Appraisal Cost |
| 4 | Requirements Rework | 6-hours | Yes | Failure Cost |
| 5 | Coding | 20-hours | No | Not Applicable |
| 6 | Code Review | 6-hours | Yes | Appraisal Cost |
| 7 | Code Rework | 2-hours | Yes | Failure Cost |
| 8 | Testing | 10-hours | Yes | Appraisal Cost |
| 9 | Test Rework | 5-hours | Yes | Failure Cost |
| 10 | Implementation | 18-hours | No | Not Applicable |

From the analysis of this time-sheet data, we can easily compute the Cost of Quality as follows:

✓ Prevention Cost = Cost of Activity at Sr. No. (1) = 10-hours

✓ Appraisal Cost = Cost of Activities at Sr. No. {(3)+(6)+(8)} = 21-hours

✓ Failure Cost = Cost of Activities at Sr. No. {(4)+(7)+(9)} = 13-hours

✓ CoQ = Cost of (Prevention+Appraisal+Failure) = 44-hours

## 6.3 Exhibit-2: *Another example of Cost Components Data Sample and CoQ-Measurement*

Let us take another example of **cost data reported**, and how it relates to CoQ-components:

| WBS-Element | Cost Incurred (in US$) |
|---|---|
| Design Reviews | $ 60,000/- |
| Tests & Inspections | $ 40,000/- |
| Excess stock (inventory) | $ 30,000/- |
| Reworks & in-house scrap | $ 20, 000/- |
| Customer returns | $ 10, 000/- |

On analyzing the above reported data sample(in-line with Exhibit-1), we can conclude that:

Total CoQ      → $ 1,30,000 ( all the above elements are part of CoQ except excess stock/inventory cost)

External Failure Cost           → $ 10, 000  (customer returns)

Internal Failure Cost           → $ 20, 000  (Reworks & in-house scrap)

Appraisal Cost           → $ 40, 000 (Tests & Inspections)

Prevention Cost           → $ 60,000  (Design reviews)

It is not the measurement, but the analysis and comparison for monitoring, control and strategic decisions that we can use the measured CoQ. Applying the concepts of CoQ measurement, analysis and corrections consistently to the SDLC-projects can help reduce the cost of quality.

## 7.   Discussion on our Research Findings

⇨    Our key findings from examination of real world SDLC-projects data selected for this study and relevant observations from series of discussions with key persons of the concerned organizations are as follows:

✍    SDLC-projects of small firms did not have any quality budget. Also, no evidence of measuring cost of quality was available on the projects of small firms.

✍    SMBs  are not following structured quality management processes. Their prevention cost is found to be 10%-15% of the total Cost of Quality.

✍    Small firms (SMBs) are only found to follow ad-hoc methodologies for quality assurance and the organization is not having any realistic idea of how much profit they are losing  thru poor quality.

✍    In big and established companies – almost all the projects claim that quality is their top priority, but only a fraction of the projects were able to provide evidence of effectively tracking cost of quality on their SDLC-projects convincingly.

✍    Even though quality was widely acknowledged as a key competitive edge, there seemed quite a few disconnect and lack of vision/commitment in various stakeholders among the top management – about quality philosophy.

✍   Average CoPQ(Cost of Poor Quality) for this study undertaken is found to be  27% of the software sales.

✍   Barring a couple of worst-case projects (where it was as high as 10-times), CoPQ is found to be 4.8 times the profits earned from the projects.

✍   In projects performing at CMMi level-5, prevention costs are found to be as high as 65% of  the total CoQ. Appraisal costs, in these cases are found to be around 20% of the total CoQ.

⇨   Another important finding of this study pertaining to **open research issues** on the subject is that software quality cost research so far has primarily been carried out by means of model building and theory generation, which has helped to develop good understanding of this research domain's structure, but  extensive empirical validation is still lacking. This lack of sufficient empirical results calls for more industry case studies having real world SDLC-projects data, because CoQ research which relies on quantitative data to generate new findings. Consequently, there is a need for more industry oriented research to gather quality cost data. This is possible with stronger cooperation between industry and academia, particularly in India.

## 8.  Conclusions:

✓   An important conclusion drawn from this study is that understanding of the cost of quality of software products, produced by SDCL-projects is extremely important in establishing a organizational quality management strategy. To arrive at the cost of quality of software products, the key considerations are:

✍   We should focus at the cost of quality over the *entire lifecycle of the software release management* – rather than just the *development life cycle*. ( i.e. considering overall software development and maintenance cost – not just the development cost alone).

✍   Once the organizational data bank of projects is established, we can compare the cost of quality with industry benchmarks and norms and plan for further improvements. No point is measuring against industry benchmarks – if organizational data bank is not in place.

✍   Identify the hidden costs related to poor quality on previous projects and document the lessons learnt. Make these documented lessons learnt - visible to all future projects. Final quality model chosen for projects should be based economic trade-offs involved with software quality.

✓   Lessons learnt from this study are articulated offer valuable prescriptive guidance for SMBs who can benefit from this study and apply the lessons learnt for quality improvement programs using CoQ-metric, to improve the capability and maturity of their SDLC-project performance, with cost effective quality of deliverables from SDLC-projects.

## 9.  Future Works

This research is a step towards further understanding of the cost of quality components in software development projects. This research has focused on possibility of using CoQ as a metric to measure the effective software quality assurance of SDLC-projects, taking the ideas from manufacturing applications.

Planned extensions of the work from this study could be:

(i)   Industry oriented Research, Case-study Analysis (from real world projects data), Understanding and Publishing of why many CoQ programs initiated with good intent have failed,  got terminated prematurely or did not deliver expected results despite investments in CoQ programs by the industry.

(ii)  Addressing concerns like quality costs do not readily appear in accounting journals.

(iii)  Taking-up studies to address  industry concerns on CoQ-programs that prevention and appraisal costs have to be stepped-up and sustained for quite a long-time, before any meaningful drop in overall CoQ could be visible on the ground (i.e. empirical research and benchmarking of cycle time between investments in CoQ programs and benefits realization).

## 10.  Limitations of Our Study

It is well known that every research/study has some limitations pertaining to methodology and dataset. In this particular case, a limitation of our study was the relatively small and convenient sample size. For this reason, our findings cannot be generalized to the broader community based on this study alone. Further empirical studies with larger data samples are needed for generalization.

## Acknowledgements

Authors would like to sincerely thank  all the PMO staff and senior project managers for sparing their time and providing their valuable inputs as a part of the semi-structured interviews that were conducted and iterated to weed-out the results. Many thanks to all the peers, colleagues and organizations who supported this research by sharing their valuable time, feedback, project data and experiences on CoQ-programs.

## References:

[1.] Webster's Online Dictionary ( http://www.websters-online-dictionary.org/definitions/quality  ).

[2.]  Crosby, Philip B.  Quality is Free: The Art of Making Quality Certain, New York: McGraw-Hill, c1979, ( ISBN 0-07-014512-1)

[3.] Crosby, P.B., *Quality Without Tears*, McGraw-Hill, New York, 1988.

[4.] Juran, Joseph and A. Blanton Godfrey. *Juran's Quality Handbook*. McGraw Hill, New York. 1999.

[5.] Deming, W. Edwards, *Out of the Crisis* (*Ch- 6: "Quality and the Consumer")*, MIT Press, 1986, ISBN: 0911570010.

[6.] A guide to the Project Management Body of Knowledge, (*Ch- 8: "Project Quality Management")*, Fourth Edition,  from PMI – World's Leading Professional Association for Project Management Profession
 ( http://www.pmi.org )

[7.] W.J. Cristopher, The cost of errors in software development: evidence from industry, The Journal of System and Software 62 (1), 2002,  pp. 1–9.

[8] Nasib S. Gill, "*Factors Affecting Effective Software Quality Management Revisited*". ACM SIGSOFT Software Engineering Notes. Volume 30, Issue 2, March 2005. Page(s): 1 – 4.

[9] Rosenberg L H, Gallo A M, Jr, "*Software Quality Assurance Engineering at NASA*". In Proceeding Aerospace Conference. Volume 5 2002 IEEE, Page(s):5-2569 – 5-2575.

[10] M Agrawal & K Chari, "*Software Effort, Quality, and Cycle Time: A Study of CMM Level 5 Projects*" *IEEE Transactions on Software Engineering*,  2007, Page(s): 145 – 156, Volume: 33, issue: 3,
    Digital Object Identifier: 10.1109/TSE.2007.29

[11] Manar Abu Talib, Adel Khelifi,  Alain Abran & Olga Ormandjieva,   "*Techniques for Quantitative Analysis of Software Quality throughout the SDLC: The SWEBOK Guide Coverage*" IEEE Proc. 8th-ACIS International Conference on Software Engineering Research, Management and Applications (SERA),  May 24-26 2010, pp. 321 - 328.

[12] A. Khanjani, &  R. Sulaiman,  "*The process of quality assurance under open source software development*" Proc. Of IEEE Symposium on Computers & Informatics (ISCI), March 20-23, 2011, pp.548-552. Digital Object Identifier: 10.1109/ISCI.2011.5958975

[13] M.K. Omar , S. Murugan, "*Investigating the impact of quality control decisions on cost of quality*", Proc. 4th IEEE International Conference on Modeling, Simulation and Applied Optimization (ICMSAO), April 19 – 21, 2011, pp. 1-6, DOI: 10.1109/ICMSAO.2011.5775569.

[14] C. Woody, N. Mead, & D. Shoemaker, "*Foundations for Software Assurance*" IEEE Proc. of 45th Hawaii International Conference on System Science (HICSS), January 4-7, 2012, pp. 5368-5374.
        Digital Object Identifier : 10.1109/HICSS.2012.287

[15] A.Janus, A. Schmietendorf, R. Dumke, J. Jager, "*The 3C approach for Agile Quality Assurance*" Proc. of 3rd International Workshop on Emerging Trends in Software Metrics (WETSoM), 3-3 June, 2012, pp 9 – 13, Digital Object Identifier : 10.1109/WETSoM.2012.6226998

[16] Wei Li, "*QoS Assurance for Dynamic Reconfiguration of Component-Based Software Systems*", IEEE Transactions on Software Engineering, pp.658 – 676, Volume: 38, issue: 3, May-June 2012. Digital Object Identifier : 10.1109/TSE.2011.37

[17] Deloitte report 2009–"Global economic slowdown and its impact on the Indian IT industry, viewed on-line at http://www.deloittemeet.com/files/IT%20Recession.pdf , July 2010.

[18] CMMI-DEV, V1.3 : CMMI® for Development, Version 1.3, a technical report on improving processes for developing better products and services from SEI – The Software Engineering Institute, Carnegie Mellon University ( http://www.sei.cmu.edu ), Pittsburgh, PA 15213, Nov 2010.

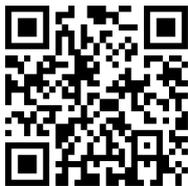

Free download this article
and more information

15